\documentclass[prb,10pt,amssymb,twocolumn]{revtex4}
\usepackage{graphicx}
\begin{document}
\title{Spectral and Diffusive Properties of Silver-Mean Quasicrystals in $1,2,$
and $3$ Dimensions}

\author{ V.~Z.~Cerovski }
\author{ M.~Schreiber }
\affiliation{ Institut f\"ur Physik, Technische Universit\"at, D-09107
Chemnitz, Germany.  } 
\author{U.~Grimm} \affiliation{ The Open University, Applied Mathematics Dept.,
Milton Keynes, MK7 6AA, UK.  }
\date{July 15, 2004}
\begin{abstract} 
Spectral properties and anomalous diffusion in the silver-mean
(octonacci) quasicrystals in $d=1,2,3$ are investigated using numerical
simulations of the return probability $C(t)$ and the width of the wave packet
$w(t)$ for various values of the hopping strength $v$.  In all dimensions we
find $C(t)\sim t^{-\delta}$, with results suggesting a crossover from
$\delta<1$ to $\delta=1$ when $v$ is varied in $d=2,3$, which is compatible
with the change of the spectral measure from singular continuous to absolute
continuous; and we find $w(t)\sim t^{\beta}$ with $0<\beta(v)<1$ corresponding
to anomalous diffusion.  Results strongly suggest that $\beta(v)$ is
independent of $d$. The scaling of the inverse participation ratio suggests
that states remain delocalized even for very small hopping amplitude $v$.  A
study of the dynamics of initially localized wavepackets in large
three-dimensional quasiperiodic structures furthermore reveals that wavepackets
composed of eigenstates from an interval around the  band edge diffuse faster
than those composed of eigenstates from an interval of the band-center states:
while the former diffuse anomalously, the latter appear to diffuse slower than
any power law.  
\end{abstract}


\maketitle

\section{Introduction}\label{sec:introduction}

The discovery of quasicrystalline solid-state structures, characterized by the
presence of long-range order but no translational symmetry \cite{Shechtman84},
which can be viewed as intermediates between crystalline and amorphous
structures \cite{Ishimasa85} with icosahedral \cite{Shechtman84}, dodecagonal
\cite{Ishimasa85}, decagonal \cite{Bendersky85}, and octagonal\cite{Wang87}
orientational symmetry, as well as the classical wave propagation in
quasiperiodic structures \cite{Torres03}, renewed the interest in studying
quasiperiodic systems, in particular their transport properties
\cite{Poon92,Poon93,Roche97,Schulz98,Mayou00}.

Quantities commonly studied to characterize the electron dynamics in this
context are the return probability $C(t)$ and the mean-square
displacement $w(t)$, defined as
\begin{equation}\label{eq:C(t)}
C(t)\equiv{1\over t}\int_0^t dt' |\Psi(\vec{r}_0,t')|^2,
\end{equation}
and
\begin{equation}\label{eq:w(t)}
w^2(t)\equiv\sum_{n=1}^{N} |\vec{r}_n-\vec{r}_0|^2 |\Psi(\vec{r}_n,t)|^2,
\end{equation}
where $\Psi(\vec{r},t)$ is a wave packet initially located at a site
$\vec{r}_0$: 
\begin{equation}\label{eq:deltawave}
  \Psi(\vec{r},t=0)=\delta_{\vec{r},\vec{r}_{0}}.
\end{equation}

One of the characteristic properties of quasiperiodic systems are power-law
asymptotic behaviors of these two quantities in the limit $t\to\infty$, when
\begin{equation}\label{eq:delta}
C(t)\sim t^{-\delta},
\end{equation}
and 
\begin{equation}\label{eq:beta}
w(t)\sim t^{\beta}.
\end{equation}
The exponents satisfy $0\le\beta,\delta\le 1$, where the two equalities
correspond to the limiting cases of, respectively, the absence of diffusion and
the ballistic motion. Inbetween these two cases anomalous diffusion takes place
(the classical diffusion is a special case $\beta=1/2$), leading to the
anomalous transport {\it via} Einstein's relation for the zero-frequency
conductivity: $\sigma\simeq 2e^2 n_F D(\tau)$, where $D(\tau)= w^2(\tau)/\tau$
is the diffusion constant, and Eq.~(\ref{eq:beta}) leads to the generalized
Drude formula $\sigma\simeq 2e^2 n_F A \tau ^{2\beta-1}$, with $\tau$ being a
characteristic time beyond which the propagation becomes diffusive due to the
scattering \cite{Roche97,Mayou00}.  Furthermore, the shape of the diffusion
front of the wavepacket has asymptotically the form of a stretched exponential
$\exp|x/w(t)|^{-\gamma}$ with $\gamma={1/(1-\beta)}$ in the limit $t\to\infty$
\cite{Zhong01}.  

The exponent $\delta$ characterizes the spectral properties and equals the
correlation dimension of the spectral measure (the local density of states)
\cite{LDOScorrdim,Zhong95b}.  Furthermore, the wavepacket dynamics exhibits
multiscaling, where different powers of displacement in Eq.~(\ref{eq:w(t)})
scale with different exponents
$\beta$~[\onlinecite{Guarneri93,Evangelou93,Guarneri94,Wilkinson94,
Piechon96,Ketzmerick97,Barbaroux01}].  Pure-point, singular continuous and
absolute continuous spectra imply, respectively, $\delta=0$, $0<\delta<1$ and
$\delta=1$, but the reverse is not necessarily true in general:
$\delta>0,\beta>0$ characterize the continuous spectra, while $\delta=0$
usually means a pure-point spectrum and the absence of diffusion ($\beta=0$),
but with nontrivial exceptions~\cite{Rio95}.  One-dimensional quasiperiodic
systems commonly have singular continuous spectra
\cite{Tang86,Zhong95b,Kohmoto87,Fujiwara89,Sire89,Bellissard89},  and may
furthermore exhibit transitions from pure-point to absolute continuous behavior
(being singular continuous at the transition point) as a parameter in the
Hamiltonian is varied, for instance in Harper's model of an electron in a
magnetic field \cite{HarpersModel}, or the kicked rotator~\cite{kickedrotator}.  

Eigenstates of quasiperiodic systems exhibit multifractality, when the electron
wavefunction is neither localized nor extended but exhibits self-similarity
\cite{Tang86,Kohmoto87,Fujiwara89,Niu86} (definitions and further discussion
are given in Sec.~\ref{sec:pr}), similar to the Anderson model of localization
for disordered conductors near the localization-delocalization transition
\cite{Schreiber91}.

The Hamiltonian $H$ studied in this work (defined in
Sec.~\ref{sec:definitions}) describes a system of $N_m^d$ atoms of a
$d$-dimensional product of $m$th approximants of the silver-mean (octonacci)
off-diagonal model.  

In a previous work \cite{Zhong95b}, the direct sum of one-dimensional
quasiperiodic systems was studied, and it was shown that already in this case
$C(t)$ displays non-trivial dimensionally-dependent behavior.  On the other
hand, it is easy to prove analytically exactly that in quasiperiodic systems
defined as a direct sum of one-dimensional quasiperiodic systems, $w(t)$ does
not depend on the dimensionality of such tilings.  In the product systems
studied here this is not the case anymore and therefore they can be viewed as a
natural next choice for the construction of interesting higher-dimensional
quasiperiodic systems in the study of $w(t)$.

The rest of the paper is organized as follows: In Sec.~\ref{sec:definitions} we
define the systems studied, whose eigenproblem is discussed in
Sec.~\ref{sec:eigenproblem}; in Secs.~\ref{sec:C(t)} and \ref{sec:w(t)} results
of the calculation of $C(t)$ and $w(t)$ are presented and discussed,
respectively; in Sec.~\ref{sec:pr} participation ratios of eigenstates are
analyzed, and finally, conclusions are presented in Sec.~\ref{sec:conclusions}.

\section{Definitions}\label{sec:definitions}

The silver-mean chain is defined over an alphabet $\{L,S\}$ by the inflation
rule 
\begin{equation}
   \rho\; :\; \left\{ \; \begin{array}{l} 
                              L  \to  LSL \\
	                      S  \to  L 
	                 \end{array} \right. ,
\end{equation}
iterated $m$ times starting with letter $S$, with the corresponding Hamiltonian
\begin{equation}\label{H_m}
H_m = \sum_{\langle n n'\rangle} t_{n n'}(c_n^\dagger c_{n'} + c_{n'}^\dagger
c_n), 
\end{equation}
where the sum is restricted to nearest neighbors and the hopping integrals
$t_{n n'}$ take values $1$ and $0\le v\le 1$ for, respectively, letters $L$ and
$S$ (``large'' and ``small'' hoppings) of the letter sequence of the
approximant (with open boundary conditions).  Among various number-theoretical
properties of the approximants, we mention that the $m$th approximant's length
$N_m$ satisfies $\lim_{m\to\infty}N_{m+1}/N_m = 1+{\sqrt 2} = \delta_S$, the
so-called silver mean.

The $d$-dimensional quasiperiodic tilings we study are derived from the direct
product of silver-mean chains~\cite{Sire89,Yuan00} with the Hamiltonian
\begin{equation}\label{eq:Hamiltonian}
H^{(d)}_m\equiv\otimes_{i=1}^d H_m.
\end{equation}
This has a particularly simple geometrical interpretation, since it describes a
particle on a $d$-dimensional cube, with coordinates
$\vec{r}\equiv\{x_1,\dots,x_d\}$, hopping only along the main diagonals of the
cube: $\{x_1,\dots,x_d\}\to\{x_1\pm 1,\dots,x_d\pm 1\}$.  Since the nearest
neighbors of the cube cannot be connected by any number of such hops from
$H_m^{(d)}$, the Hamiltonian~(\ref{eq:Hamiltonian}) decomposes into $N=2^{d-1}$
parts defined on corresponding interpenetrating disconnected quasiperiodic
tilings.  It can be shown that all of these Hamiltonian parts are equivalent,
and that the tilings they define are symmetry related, due to the reflection
symmetry of the silver-mean chain about the middle bond of the chain.  

\begin{figure}
\begin{center}
\includegraphics*[width=3.0in]{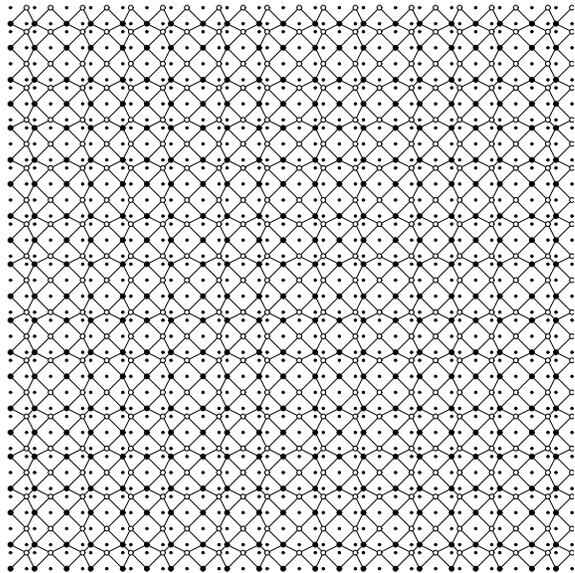}
\end{center}
\caption{Labyrinth tiling corresponding to the
Hamiltonian~(\ref{eq:Hamiltonian}) for $m=5, d=2$.  The distances $L$ and $S$
of the silver-mean chain are taken as $1$ and $v=0.5$.  Then the bond lengths
of the labyrinth tiling are $\sqrt{2}, \sqrt{1+v^2},$ and $v\sqrt{2}$.  The
bipartite structure is indicated by the open and closed circles for the sites,
taking into account that hopping occurs only among the sites of different kind.
The second, equivalent, labyrinth tiling is shown in the figure by dots.}
\label{fig:labyrinth}
\end{figure}

Such tilings in $d=2$ are known as labyrinth tilings, and
Figure~\ref{fig:labyrinth} shows one example, an $m=5$ tiling.  The tiling has
bipartite structure.  Furthermore, a two-dimensional projection of the
silver-mean tiling in $d=3$, for instance, looks exactly as the tiling in
Figure~\ref{fig:labyrinth}, because the whole tiling has a layered structure:
all symbols of the same kind in the figure belong to one layer.  The distance
between successive layers are, of course, again determined by the silver-mean
sequence.  The number of sites $N$ in various tilings is given in the
Table~\ref{table:Nm}.

\begin{table}
\begin{center}
\begin{tabular}{r|r r r}
\hline
$m$ & $d=1$ & $d=2$ & $d=3$ \\
\hline
 $2$ & $    4$ & $         8$ & $            16$ \\
 $3$ & $    8$ & $        32$ & $           128$ \\
 $4$ & $   18$ & $       162$ & $          1458$ \\
 $5$ & $   42$ & $       882$ & $         18522$ \\
 $6$ & $  100$ & $      5000$ & $        250000$ \\
 $7$ & $  240$ & $     28800$ & $       3456000$ \\
 $8$ & $  578$ & $    167042$ & $      48275138$ \\
 $9$ & $ 1394$ & $    971618$ & $     677217746$ \\
$10$ & $ 3364$ & $   5658248$ & $    9517173136$ \\
$11$ & $ 8120$ & $  32967200$ & $  133846832000$ \\
$12$ & $19602$ & $ 192119202$ & $ 1882960298802$ \\
\hline
\end{tabular}
\end{center}
\caption{The number of sites $N$ of the $d$-dimensional tilings of various
approximants $m$ of the silver-mean (octonacci) quasicrystal. }
\label{table:Nm}
\end{table}

\section{Eigenproblem}\label{sec:eigenproblem}

The eigenproblem of $H_m$ is
\begin{equation}\label{eq:1d}
  H_{m}|\psi_i\rangle = E_i|\psi_i\rangle\;, \;\;\;\;\; i = 1,\dots, N_m,
\end{equation}
to which the eigenproblem of $H_m^{(d)}$ reduces:
\begin{eqnarray}\label{eq:H_m^deigenproblem}
  H_{m}^{(d)}|\psi_{i_1}\rangle\otimes\cdots\otimes|\psi_{i_d}\rangle 
  &\equiv& E_{i_1}\cdots
  E_{i_d}|\psi_{i_1}\rangle\otimes\cdots\otimes|\psi_{i_d}\rangle\;,\cr 
  & &\;\;\;\; i_j = 0,\dots,N_m.
\end{eqnarray}

Regarding the subtiling eigenproblem, its eigenstates are obtained simply by
projecting the eigenstates of $H_m^{(d)}$ onto the subtiling (with
normalization constant ${N}^{-1/2}$) due to the equivalence of the individual
eigenproblems~(\ref{eq:1d}).  Thus, if $P$ denotes the projector onto the
subspace of one subtiling, the corresponding subtiling Hamiltonian is
$PH^{(d)}_mP$ and its eigenstates are
${N}^{-1/2}P|\psi_{i_1}\rangle\otimes\cdots\otimes|\psi_{i_d}\rangle$ with the
same eigenenergies as in Eq.~(\ref{eq:H_m^deigenproblem}).  The reduction of
the problem to quasiperiodic subtilings of the product tiling therefore only
reduces the degeneracy of individual eigenstates of $H^{(d)}_m$, while the set
of distinct eigenenergies is exactly the same.  Furthermore, since all the
Hamiltonians studied here have only off-diagonal matrix elements, their
eigenenergies necessarily come in opposite-energy pairs, and thus we can
restrict our discussion to the spectral properties for states with $E\le 0$.

\begin{figure*}
\begin{center}
\includegraphics*[height=5.4in,bb=0 0 593 640,draft=true]{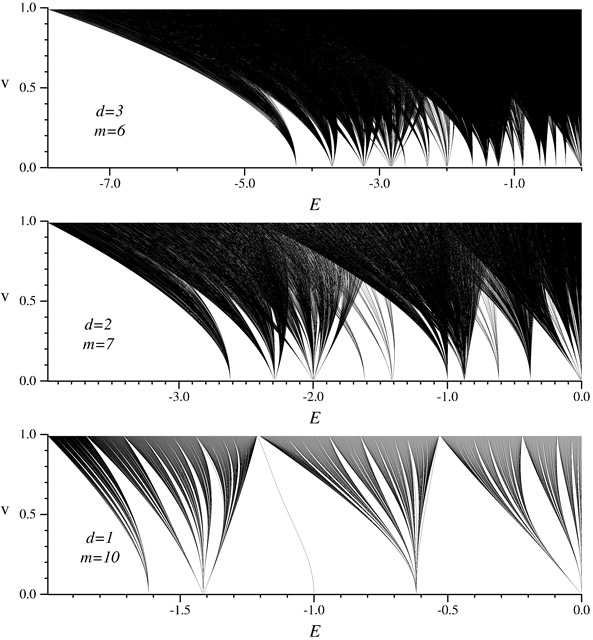}
\end{center}
\caption{Dependence of eigenenergies on the parameter $v$ in $d=1,2,3$, from
the bottom panel to the top, respectively, for the iterants $m=10,7,6$.  Only
eigenenergies $E<0$ are shown due to the symmetry of the spectra about the band
center. 
}~\label{fig:E(v)}
\end{figure*}

Figure~\ref{fig:E(v)} shows the dependence of the eigenenergies on the
parameter $v$.  A characteristic feature of the spectra is the existence of
gaps of various sizes.  In $d=1$, the gap-labeling theorem gives an enumeration
of all the possible gaps in the spectrum \cite{Kohmoto87,gaplabeling}.  For
$d>1$ these gaps seem to close for a sufficiently large $v$, and
Fig.~\ref{fig:E(v)} suggests that gaps persist only for $v\lesssim 0.6, 0.4$ in
$d=2,3$, respectively.  

For $d>1$ the gaps close for smaller values of $v$ due to the level crossings
that are a consequence of the factorizability of the eigenenergies in
Eq.~(\ref{eq:H_m^deigenproblem}).  Due to this we expect that the spectrum will
acquire finite Lebesgue measure and change from from fractal to continuous for
intermediate values of $v$, similar to the results of
Ref.~[\onlinecite{Sire89}].

A more difficult question, that we address in the next section, is whether the
spectrum is singular or absolute continuous or a mixture of the two in
different regions of the spectrum for various values of $v$.  To this end we
first note that $2^d$ states corresponding to $E=-1$ for $v=0$ are a
consequence of the open boundary conditions used in this work, corresponding to
the electron localized, for small $v$, at the ends of the chain (in $d=1$ and
corners of the cube for $d>1$), and spreading across the system as $v\to 1$.
In this respect the systems studied here are different from those of
Ref.~[\onlinecite{Yuan00}], where two $E=0$ states have been induced by adding
two sites at each end of the chain with $\psi=0$.  However, for the results
presented here this difference in boundary conditions does not play any
important role.

\section{Return probability and scaling of the spectral measure}\label{sec:C(t)}

\begin{figure*}
\begin{center}
\includegraphics*[width=6.5in]{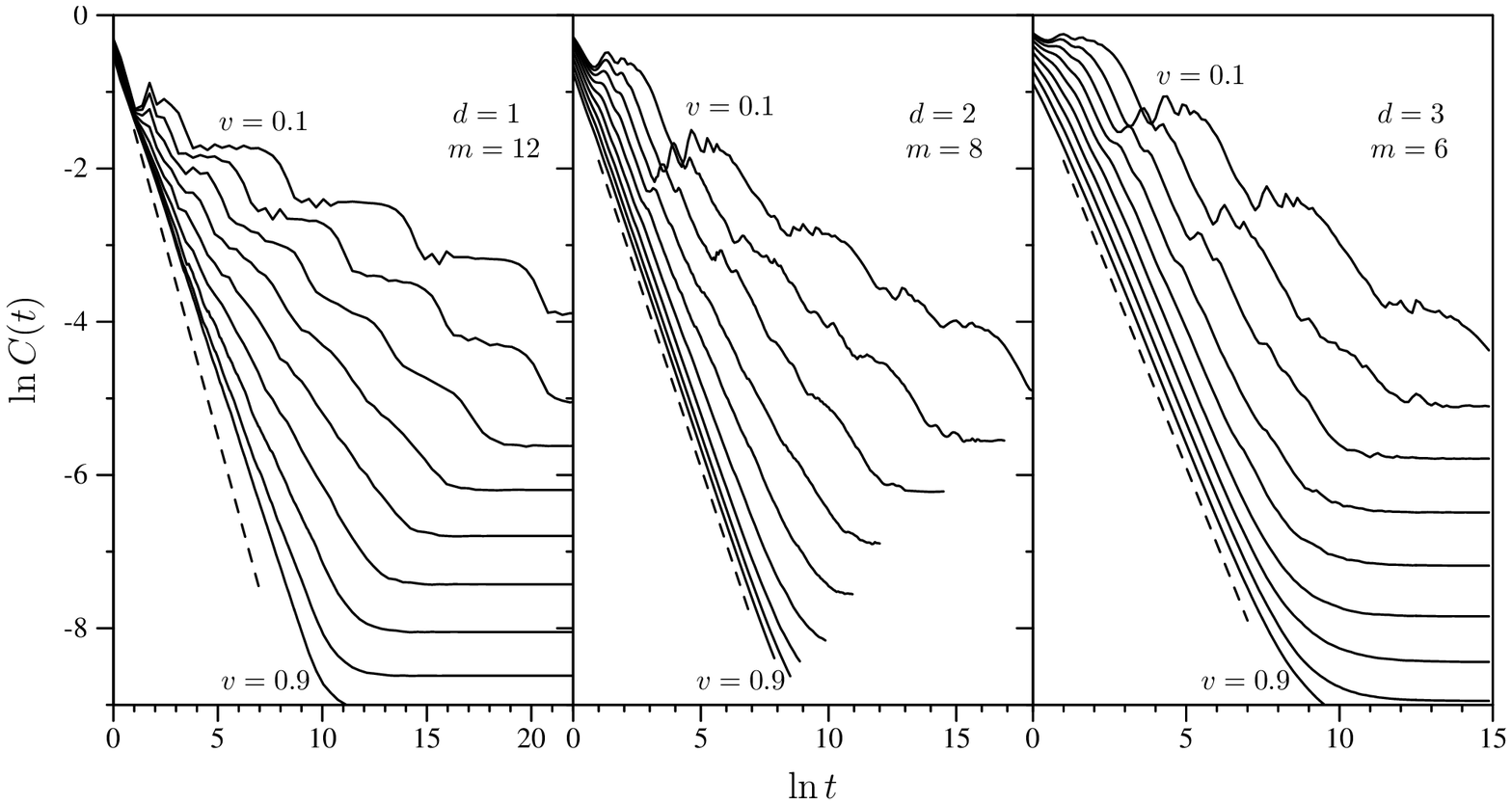}
\end{center}
\caption{Calculated values of the return probability $C(t)$ in $d=1,2,$ and
$3$, from left panel to the right, respectively, for $v=0.1, 0.2, \dots, 0.9$,
respectively, from the top curve to the bottom. The dashed line in each panel
has the slope $-1$.} \label{fig:C(t)} 
\end{figure*}

In this section we study the scaling of the return probability $C(t)$ by
evaluating numerically Eq.~(\ref{eq:C(t)}) for the eigenstates of a silver-mean
tiling in $d=1,2,3$ constructed, as described in the previous section, from
eigenstates of the chain, obtained from the numerical diagonalization of the
$H_m$, and taking into account that the time evolution is given by $\psi_j(t) =
\psi_j\exp(- i E_j t)$.  We present the results in Fig.~\ref{fig:C(t)}.  For
small values of $v$ there are pronounced steps, and this limit we analyze
elsewhere~\cite{CGS}.

For each of the studied cases in the figure, there are three characteristic
time-scales of the particle dynamics: (i) short times, characterized by a high
probability to find the particle at the initial site, which corresponds to the
regime when the wavepacket has only begun spreading; (ii) intermediate times,
when $C(t)$ does seem to behave according to the power law~(\ref{eq:delta});
and (iii) the long-time limit when there is a crossover into a constant $C(t)$,
due to the finite spatial extent of the studied systems.

The quantitative determination of $\delta$ is rather difficult, due to the
large subdominant contributions to the asymptotic power-law behavior
\cite{Zhong95b}.  These appear through the dependence of $\delta$ on the
interval of $t$ values taken for the least mean-squares linear fit in the
log-log plot of $C(t)$.  Lacking rigorous results, one attempt to characterize
these corrections to Eq.~(\ref{eq:delta}) would be to consider a class of
power-law subdominant corrections by assuming
\begin{equation}\label{eq:subdominant}
  C(t) \sim t^{-\delta} + c_1 t^{-\delta'} \;,\;\; 0<\delta\le\delta'\le 1.
\end{equation}
It was noted in Ref.~[\onlinecite{Zhong95b}] that, in the case of the Fibonacci
model, $C(t)$ can be accurately fitted to Eq.~(\ref{eq:subdominant}) under the
assumption that $\delta'$ is exactly equal to $1$, and the authors gave some
plausible but non-rigorous arguments for this assumption.  On the other hand,
numerical verification of Eq.~(\ref{eq:subdominant}) by a least-squares
nonlinear fit requires that $C(t)$ is calculated for many decades of $\ln t$.
In addition, verification of $\delta'\approx 1$ by such a procedure would be
particularly difficult near the transition where the fit would be expected to
yield already $\delta\to 1$, {\it i.e.} to decompose the fitted function into
two nearly identical terms.  Finally, the parameter $v$ would be expected to
play some role also in the subdominant terms, for instance through the
dependence of the value of $\delta'$ on $v$.

An additional difficulty was pointed out in Ref.~[\onlinecite{Oliviera99}],
where the scaling exponent of the second moment of the spectral measure was
considered (which is known~\cite{LDOScorrdim} to be equal to $\delta$) for
several quasiperiodic systems, including Fibonacci chains, and an apparently
irregular behavior of the moment was found in all of the non-Fibonacci models.
This led the authors to use a fitting procedure that does not involve fitting
of a straight line and even to consider the possibility that the exponent
$\delta$ might not exist in these non-Fibonacci systems.

In this work we fit the power law~(\ref{eq:delta}) to the calculated values of
$C(t_i)$ (for many values of $t_i$ equidistant in the logarithmic scale) by
fitting of the expression
\begin{equation}\label{eq:fit}
  C(t) = A t^{-\delta + \kappa\ln t},
\end{equation}
in all intervals containing $t_i$ values spanning at least one decade in $t$
(for easier comparison, this corresponds to a change in $\ln t$ of at least
$2.3$ in the plots), and selecting among the obtained fits of $\delta$ those
that have $\kappa$ values closest to $0$.  Should several distinct values of
$\delta$ occur that all have small $\kappa$, it seems reasonable that those
values corresponding to later times are closer to the correct asymptotic value
of $\delta$, just as it is reasonable to expect that for larger $t$ values
Eq.~(\ref{eq:subdominant}) becomes close to the power law~(\ref{eq:delta}).
Our fitting procedure does not assume anything in particular about the
subdominant terms of the powerlaw asymptotic behavior, but rather estimates
$\delta$ by requiring the absence of certain types of subdominant terms, as
quantified by $\kappa$.   

\begin{figure}
\begin{center}
\includegraphics*[width=3.5in]{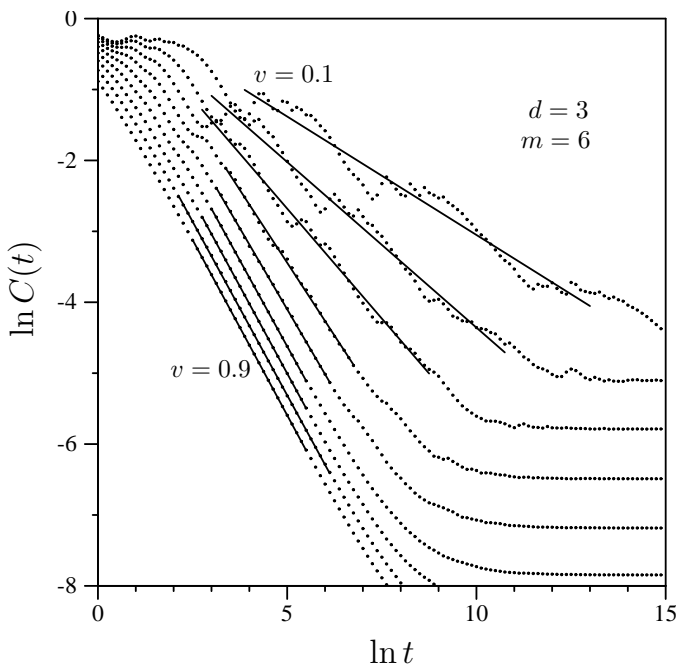}
\end{center}
\caption{Return probability in $d=3$ (points) together with fitted lines close
to the power law from the procedure described in the
text.}\label{fig:C(t)fitted}
\end{figure}

We show such obtained fits in Fig.~\ref{fig:C(t)fitted} for the case $m=6,
d=3$.  This is the most difficult system among the three cases presented in
Fig.~\ref{fig:C(t)} because the duration of the intermediate time regime (ii),
where the power law asymptotic behavior should be most likely expected, depends
only on the linear size of the system, determined by $m$, and not on the
dimensionality $d$.  The obtained values of $\kappa$ in the fits presented in
Fig.~\ref{fig:C(t)fitted} range from $10^{-4}$ to $10^{-7}$.  

A characteristic feature of the fits is that for $v\gtrsim 0.5$, the time
intervals where the power law is obeyed best ({\it i.e.} where $\kappa$ is
smallest) are shorter than the intervals for $v\lesssim 0.4$, even though for
large $v$ the curves for $\ln C(t)$ do not show oscillatory behavior and appear
to be more similar to the straight lines.  There is, however, a small curvature
in the dependence of $\ln C(t)$ on $\ln t$ for the crossover times inbetween
regimes (i) and (ii) as well as (ii) and (iii).  This is hard to see in the
plots, since it involves large relative changes in small curvatures, but easy
to distinguish numerically in our fitting procedure.

\begin{figure*}
\begin{center}
\includegraphics*[width=6.5in]{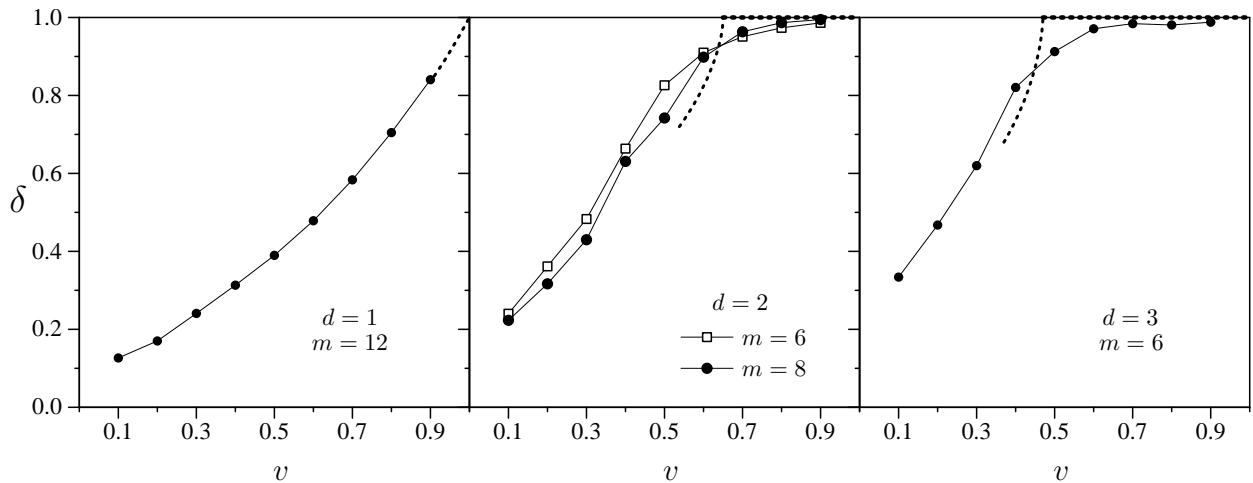}
\end{center}
\caption{Calculated values for the exponent $\delta$ characterizing the
asymptotic behavior $C(t)\sim t^{-\delta}$.  Error bars are of the size of the
symbols.  Connecting lines are included for guiding the eye.  The possible
behavior of $\delta$ in the limit of large $m$ is indicated by dotted lines
(see text for further discussion).  }\label{fig:delta}
\end{figure*}

The results for $\delta$ obtained from the above described fitting procedure
applied to the results from Fig.~\ref{fig:C(t)} are presented in
Fig.~\ref{fig:delta}.  The plot for $d=2$ shows $\delta(v)$ also for $m=6$,
which is systematically larger than $\delta(v)$ of the $m=8$ system for $v<v_c$
and smaller for $v>v_c$, from which we speculate on the possible form of the
limiting curve of the infinite system as indicated in the figure for $d=2,3$.
This extrapolation of $\delta(v)$ in $d=2,3$ is in agreement with
semi-quantitative considerations of this limit in Ref.~[\onlinecite{Zhong95b}].
For $d=1$ the results suggest that $\delta<1$ for all $v<1$ also in the
infinite system.  The exponents thus suggest that there is a singular
continuous spectrum for all $v$ values in $d=1$ and a possible transition from
singular continuous to absolute continuous spectra at $v_c\approx 0.6$ in $d=2$
(in agreement with a previous estimate~\cite{Yuan00}), and at $v_c\approx
0.4$--$0.5$ in $d=3$.  

A difficulty in obtaining exact values of $v_c$ could be due to the possibility
that, for $v\approx v_c$, the spectrum is a mixture of singular continuous and
absolute continuous parts, when absolute continuous parts appear in only small
intervals of the full spectrum that grow larger with increasing $v$, and when
the $\delta$ values obtained for $v$ away from $v_c$ are closer to the correct
asymptotic values.  

A previous study \cite{Yuan00} of the propagation of the projection of the wave
packet~(\ref{eq:deltawave}) onto various segments of the spectrum of the
silver-mean model in $d=2$ found that $C(t)$ of such restricted wave packets
still obeyed the power law, with exponents differing only slightly from
$\delta$ obtained when the full spectrum is used.  We were able to extend this
result by proving the following: If the spectrum is divided into a finite
number of $M$ segments such that $1 = \sum_{\alpha=1}^M P_\alpha$, where
$P_\alpha$ is the projector onto the subspace corresponding to the segment
$\alpha$, and if $C_\alpha(t)\sim t^{-\delta_\alpha}$ in each of the $M$
segments, where $C_\alpha(t)$ is defined as in Eq.~(\ref{eq:C(t)}) with the
wavepacket~(\ref{eq:deltawave}) replaced by its projection
$P_\alpha\Psi(\vec{r},0)$, and if Eq.~(\ref{eq:delta}) also holds, then $\delta
\ge \max_\alpha \{\delta_\alpha\}$.

\section{Anomalous diffusion}\label{sec:w(t)}

The main advantage of calculating the width $w(t)$ of a wave packet is that it
is more directly related to the transport properties than $C(t)$, as discussed
in the introduction, and that its asymptotic power-law behavior~(\ref{eq:beta})
seems to be less influenced by subdominant contributions, and therefore easier
to determine from numerical studies of finite-size systems, as we discuss
below. 

One of the main difficulties in numerical studies of the anomalous diffusion is
that in order to obtain $w(t)$ accurately, many eigenvectors are needed.
Calculating its behavior at large times also requires large system sizes.  This
puts significant constraints on the system sizes that can be investigated
numerically, and several approximations have been developed to circumvent these
difficulties.  One of them is to expand the evolution operator in terms of
small-time increments \cite{Kawarabayashi96}, or to study the time evolution of
the position operator expanded in terms of Chebyshev polynomials combined with
energy filtering {\it via} a Gaussian operator centered at a given energy and
of the width of, {\it e.g.}, $1\%$ of the total bandwidth \cite{Triozon03}.  In
the latter work, generalized quasiperiodic Rauzi tilings have been studied and
the authors were interested how the topological connectivity of the tiling
influences transport properties of the quasicrystal in $d=2,3$.  The
simplifying feature of the model is that the Hamiltonian is sparse with no
matrix elements different from $1$, which significantly speeds up any algorithm
where multiplication with the Hamiltonian is a time-consuming part of the
calculation, like the algorithm of [\onlinecite{Triozon03}] as well as, for
instance, the Lanczos algorithm.  Another approach was to study the evolution
of a wave packet $\Psi_\alpha(\vec{r},t=0)$ constructed from a subset of
eigenstates from an interval $\alpha$ of the full spectrum\cite{Yuan00},
\begin{equation}
 \Psi_\alpha(\vec{r},t=0) \propto P_\alpha \delta_{\vec{r},\vec{r_0}}.
\end{equation}
Using such wave packets, the largest system studied in this work is $m=7$ for
$d=3$ with 3456000 sites.

\begin{figure*}
\begin{center}
\includegraphics*[width=6.5in]{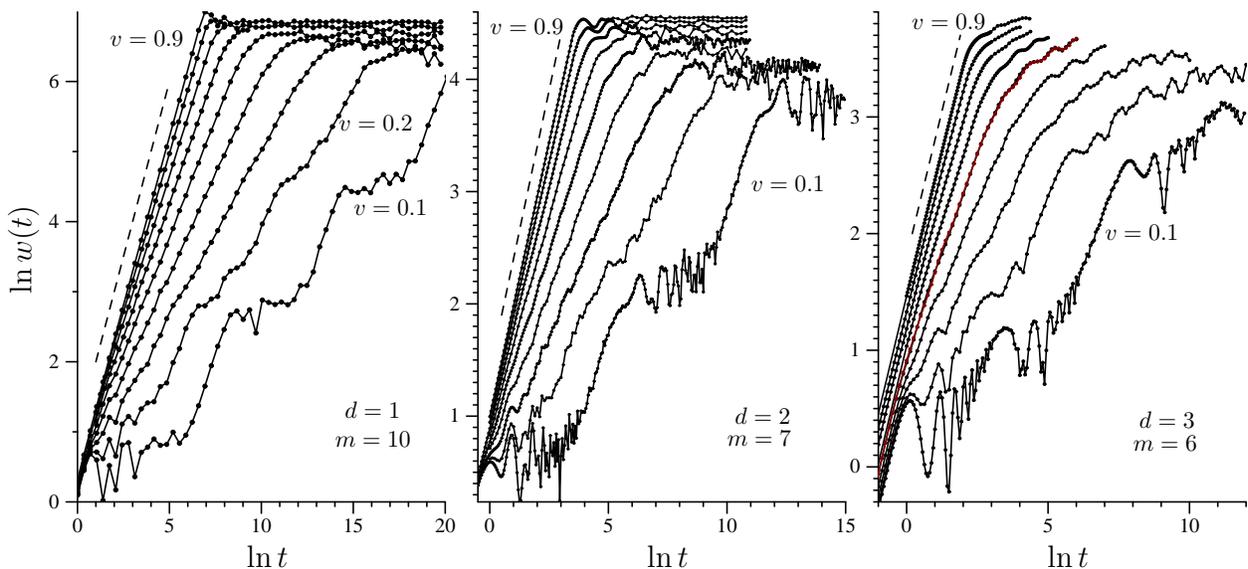}
\end{center}
\caption{Calculated values of the width $w(t)$ of the wave packet for three
systems as indicated in each panel, for $v=0.9, 0.8, \dots, 0.1$, going from
top curve to the bottom, respectively.  The dashed line in each panel has slope
1 corresponding to the ballistic motion.}\label{fig:w(t)} \end{figure*}

Figure~\ref{fig:w(t)} presents results of the calculations of the spreading
width $w(t)$ of a wave packet initially localized on a single site using all
the eigenstates of the spectrum.  For small $v$ values pronounced steps appear,
related to the similar steps in Fig.~\ref{fig:C(t)} and the structure of very
narrow bands of the spectra (cp. Fig.~\ref{fig:E(v)}), that we discuss
elsewhere \cite{CGS}.  

The three regimes of wave-packet dynamics discussed in the previous section
here correspond to (i) the nearly ballistic propagation for early times
$t\lesssim 1$, (ii) anomalous diffusion, in intermediate intervals of $t$,
characterized by Eq.~(\ref{eq:w(t)}) and (iii) a stationary regime of
approximately constant $w(t)$, due to the finite extent of the studied systems,
for large $t$.  The maximum of $w$ which would be possible for a system of
linear size $L$ can be estimated as $\sqrt{d}L/2$, corresponding to a state
completely localized at the corners of the sample; for the system sizes in
Fig.~\ref{fig:w(t)}, maximal values of $\ln w$ are approx.~$7.4, 5.1, 4.5$ in
$d=1,2,3$, respectively.  On the other hand, a completely delocalized state
would lead to $\ln w \approx 6.878, 4.587, 3.919$, respectively; thus
Fig.~\ref{fig:w(t)} shows that for large values of $v$ the wave packet becomes
smeared out over the entire sample in the regime (iii).  The crossover from
regime (ii) to (iii) becomes longer for larger $d$, which is due to the more
complicated reflection of the wave packet off the boundaries in
higher-dimensional tilings.

\begin{figure}
\begin{center}
\includegraphics*[width=3.0in]{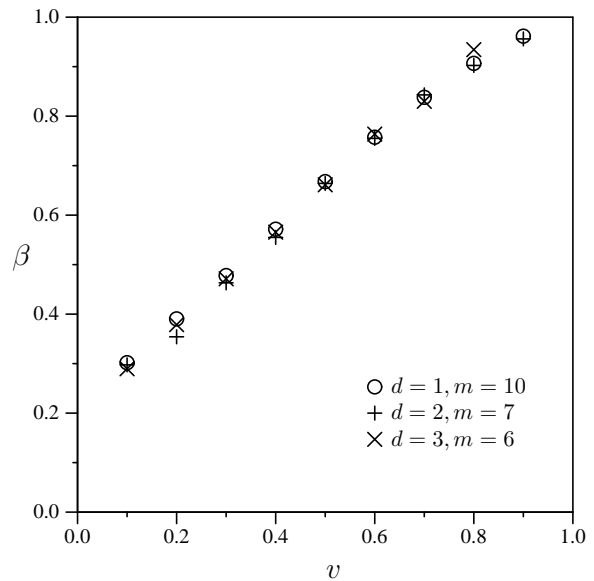}
\end{center}
\caption{Values of the exponent $\beta$ governing anomalous diffusion of the
wave packet, as extracted from the data in Fig.~\ref{fig:w(t)}.  The error bars
are of the size of the symbols. }\label{fig:beta} \end{figure}

The values of the exponent $\beta$, which were obtained using the same fitting
procedure as in the previous section, are presented in Fig.~\ref{fig:beta}.
They seem to be independent of the dimensionality of the system.  Since the
calculation of $w(t)$ is computationally slower than that of $C(t)$, system
sizes considered here are smaller than in the previous section, and in
particular, for $v=0.9$, the power law was observed for only about a half a
decade of time for $d=2, m=7$, as opposed to two decades for $d=1, m=10$, and
we could not obtain a reliable exponent using our fitting procedure for $d=3,
m=6, v=0.9$.  In the fitting procedures used in the previous section, for
comparison, the calculated exponents were obeyed for at least a decade (and up
to 6 decades in some cases for smaller $v$ values).

In order to assess the role of topology of the quasiperiodic tiling in $d=3$,
we make a comparison of the wave-packet dynamics studied here with those of
Ref.~[\onlinecite{Triozon03}].  Since in the latter work energy-filtering was
used as described above, we study the spreading of the wave
packet~(\ref{eq:deltawave}) composed only of eigenstates with energy $E$ in the
interval $0<|E|-|E_\alpha|<\Delta E_\alpha$, where $\Delta E_\alpha$ is chosen
such that the interval contains about $2\%$ of the total number of states.

\begin{figure*}
\begin{center}
\includegraphics*[width=5in]{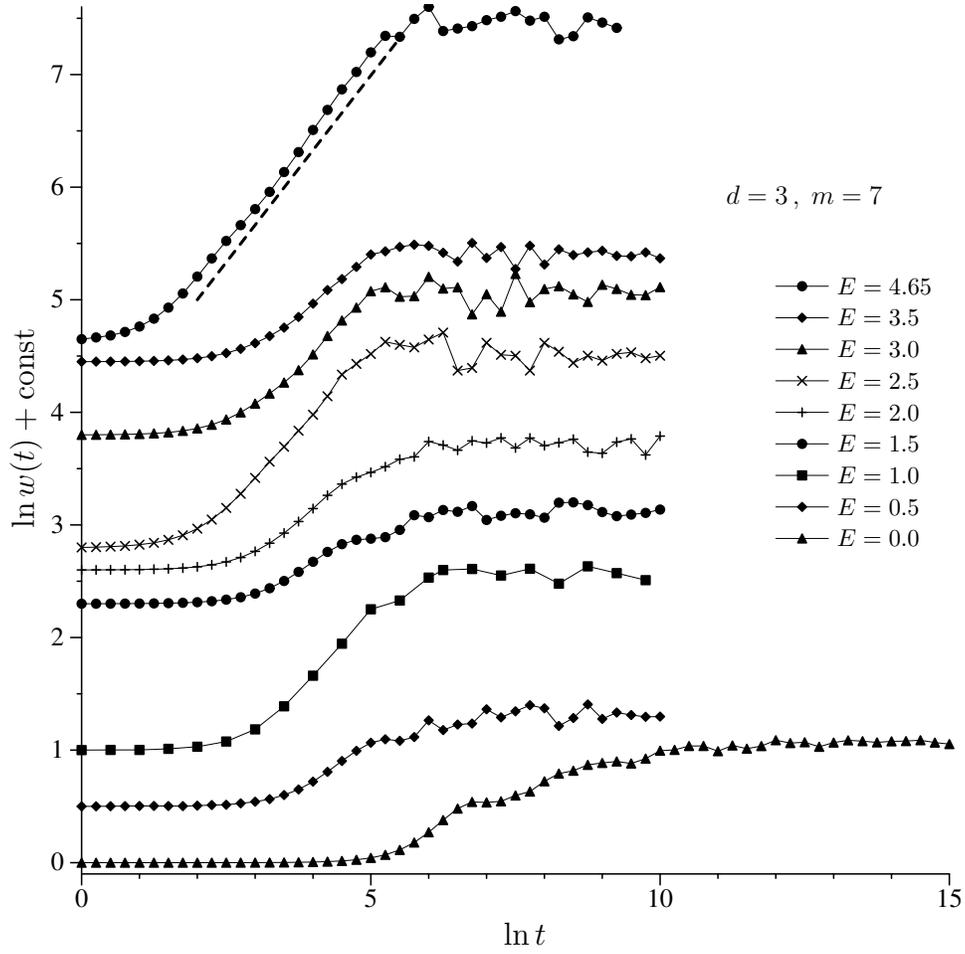}
\end{center}
\caption{Calculated values of the width $w(t)$ of the initially localized wave
packet composed only of states with eigenenergies from small energy intervals
containing around $2\%$ of the total number of states, for $d=3, m=7, v=0.5,$
(with $3456000$ sites) going from the band edges (top curve) to the band center
(bottom curve).  Individual curves are shifted vertically for clarity.  The
dashed line near the topmost curve has the slope $\beta\approx 0.66$ that was
found in the $m=6$ system using the whole spectrum.}\label{fig:d(t)part}
\end{figure*}

Figure~\ref{fig:d(t)part} presents calculated values for several energies
$E_\alpha$ between the band edge and the band center.  Apparently when the wave
packet is composed only of band-edge states, the diffusion is anomalous with
the exponent very close to the value $\beta\approx 0.66$ obtained when the full
spectrum is considered, as indicated in the figure.  For other energies, there
seems to be an intermediate time regime when anomalous diffusion also takes
place, with the same or smaller exponents as compared to the one found at the
band edge.  Furthermore, a third kind of behavior also occurs, most notably at
the band center, where the wave packet seems to spread slower than any power
law.  This is in qualitative agreement with the findings of
[\onlinecite{Triozon03}] in the sense that there wave packets made out of
states from the band edge also spread faster but, in contradistinction with the
finding here, ballistic motion was found at the band edge and anomalous
diffusion at the band center.   

It is unclear to us at present what kind of eigenstates are responsible for the
slow diffusion seen at the band center.  Qualitatively,  marginally critical
states~\cite{Fujita01} might be perhaps related to this kind of wavepacket
dynamics.

\begin{figure*}
\begin{center}
\includegraphics*[width=6in]{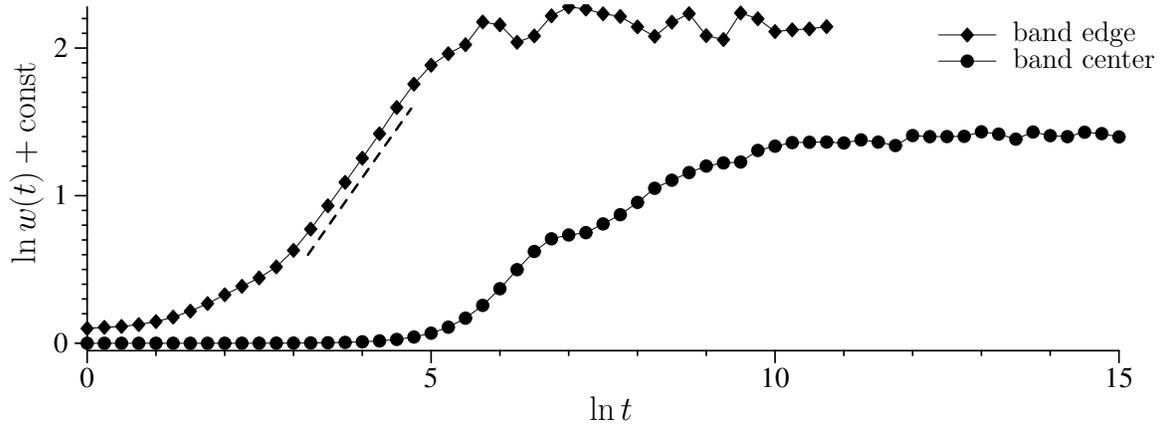}
\end{center}
\caption{Same as Fig.~\ref{fig:d(t)part} but for a slightly different initial
site of the wave packet (see text for further details).  The dashed line has
the slope $\beta\approx 0.66$, same as in Fig.~\ref{fig:d(t)part}.
}\label{fig:d(t)partb}
\end{figure*}

Such constructed wavepackets will inevitably have some ``ripples'' that depend
on how well eigenstates from the chosen interval $\alpha$ approximate
$\delta_{\vec{r},\vec{r}_0}$, which in turn depends on the choice of the
starting site $r_0$.  This can be characterized by
$p_0=|P_\alpha\delta_{\vec{r},\vec{r}_0}|^2$, and states in
Fig.~\ref{fig:d(t)part} have $p_0\approx 14\%$ and $0.7\%$ at the band edge and
the band center, respectively.  In order to show some evidence that the
obtained dynamics in the two cases does not depend on the choice of the initial
site, we present $w(t)$ in Fig.~\ref{fig:d(t)partb} when the initial site is
chosen two sites along the main diagonal of the system away from the center, in
which case $p_0\approx 0.3\%$ and $11\%$ at the band edge and the band center,
respectively.  The dependence of the values of $p_0$ on the choice of the
initial site is reflected in the range of values that $w(t)$ takes, but does
not seem to affect the two different types of dynamics.  Although the
contribution $p_0$ of the band edge states to the wave packet in
Fig.~\ref{fig:d(t)partb} is much smaller than to that in
Fig.~\ref{fig:d(t)part}, the spreading of the entire wave packet is again
characterized by the slope $\beta\approx 0.66$ in agreement with the value
shown in Fig.~\ref{fig:beta}.

\section{Participation Ratios}\label{sec:pr}

The results of the previous section strongly suggest that $\beta(v)$ is
independent of $d$ for the class of systems $H_m^{(d)}$ studied here.  In this
section we attempt to link that with the eigenstate properties by the analysis
of the participation number.

The (generalized) inverse participation numbers, $Z_q(\psi)$, are defined as
\begin{equation}
  Z_q(\psi)\equiv\sum_r |\psi(r)|^{2q}.
\end{equation}
The participation number
$
  P_2(\psi)\equiv Z_2^{-1}(\psi),
$
for instance, characterizes on how many sites a given state $\psi$ is
significantly different from zero: for a particle completely localized at a
single site $P_2(\delta_{\vec{r},\vec{r_0}})=1,$ while for a Bloch wave
$P_2(\exp(i\vec{k}\cdot\vec{r})/{\sqrt N}) = N$.  The participation ratio,
$p_2(\psi)\equiv P_2(\psi)/N$, gives the fraction of the total number of sites
where $\psi$ is significantly different from 0.

In the case of the eigenstates of $H_{m}^{(d)}$, we have
\begin{eqnarray}
  Z_q( |\psi_{i_1}\rangle\otimes\cdots\otimes|\psi_{i_d}\rangle ) & = &
  \sum_{x_1}\cdots\sum_{x_d}|\psi_{i_1}(x_1)\cdots
  \psi_{i_d}(x_d)|^{2q}\nonumber \\ & = &
  \sum_{x_1}|\psi_{i_1}(x_1)|^{2q}\cdots\sum_{x_d}
  |\psi_{i_d}(x_d)|^{2q}\nonumber\\ & = & Z_q(\psi_{i_1})\cdots
  Z_q({\psi_{i_d}}),\label{eq:productZq}
\end{eqnarray}
or, in other words, $Z_q$ of the product state is equal to the product of the
inverse participation numbers of the one-dimensional states.

On the other hand, the inequality
\begin{equation}\label{eq:inequality}
  (\min_{i} Z_q(\psi_i))^d \le Z_q(\psi_{i_1})\cdots Z_q({\psi_{i_d}}) \le
  (\max_{i} Z_q(\psi_i))^d, 
\end{equation}
gives us insight into the nature of eigenstates of $H_m^{(d)}$: they are always
less or equally extended than the most extended state of the product, and more
or equally extended than the least extended state of the product.  In
particular, they can be anisotropically extended, which happens whenever states
from the product have quite different participation numbers.  For each such
state of $H_{m}^{(d)}$, there is another one with the same energy and
anisotropy in a different direction due to the product structure of the
Hamiltonian.

\begin{figure}
\begin{center}
\includegraphics*[width=3in]{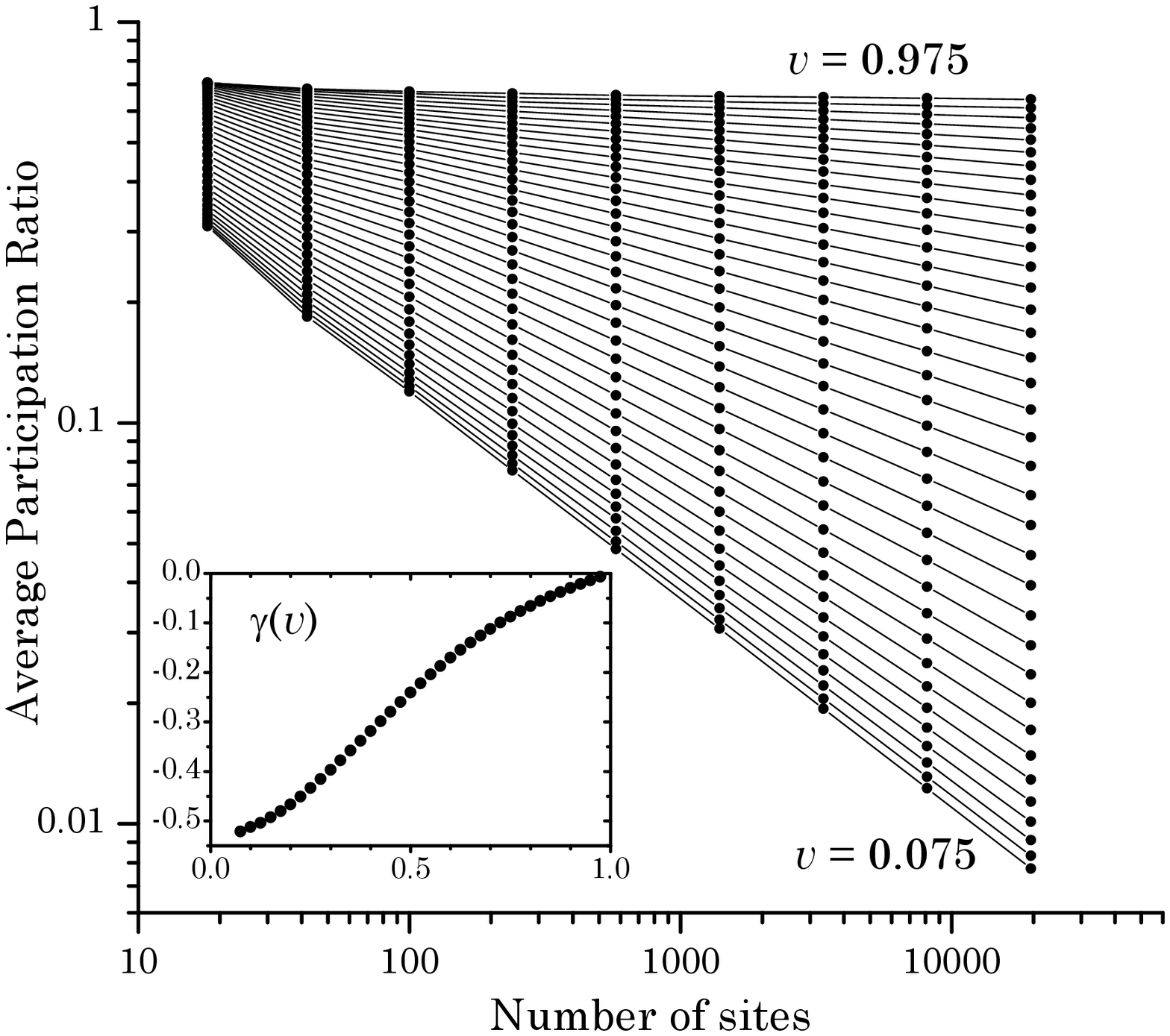}
\end{center}
\caption{Dependence of the average participation ratio $p_2(v)$ on the number
of sites $N_m$ of the silver-mean chain for $m=4,\dots,12$ in a
doubly-logarithmic plot for $v=0.075,0.1,0.125,\dots,0.975$. The inset shows
the values of the exponent $\gamma(v)$ obtained from the fit of $p\sim
N^\gamma$.}\label{fig:pr}
\end{figure}

A previous numerical study \cite{Yuan00} indicates that the scaling exponents
of $Z_2$ of the ground state and the band center state are only slightly
different.  Here, we investigate the scaling of the participation ratio $p$
averaged over the whole spectrum as a function of $N_m$ and $v$ in $d=1$, and
the results are presented in Fig.~\ref{fig:pr}.

The results indicate that the average participation ratio scales with a power
law in the number of sites of the chain,
\begin{equation}\label{eq:gamma}
   p_m(v) \sim N_m^\gamma(v),
\end{equation}
and the inset of Fig.~\ref{fig:pr} shows the values of $\gamma(v)$ obtained
from the least mean-squares linear fit to the data for the largest system sizes
in the figure.  The results suggest that $\gamma(v\to 1)=0$, which is not
surprising since in this limit the quasiperiodic tiling becomes a periodic
chain (we note that in our case $p(v\to 1)< 1$ because we have real instead of
complex eigenstates so that the Bloch waves have nodal structure).  On the
other end, however, $\gamma(v\to 0) =-0.52\pm 0.03$, even though for $v=0$ the
chain breaks up into clusters of atoms, and therefore $\gamma(v=0)=-1$.  This
particular limit we address elsewhere \cite{CGS}.   For $d>1$ we have also
calculated $\gamma(v)$ numerically for several $v$ values and it remains the
same within the error bars, which we attribute to the multiplicative nature of
the participation numbers as reflected in Eqs.~(\ref{eq:productZq})
and~(\ref{eq:inequality}).  This can be related to the independence of
$\beta(v)$ with respect to the dimensionality determined in the previous
section.

\section{Conclusions}\label{sec:conclusions}

We studied spectral properties, dynamics of wave packets and scaling properties
of eigenstates in 1-, 2- and 3-dimensional systems obtained as direct products
of the silver-mean chains, by investigating the scaling exponents $\delta,
\beta$ and $\gamma$ describing the asymptotic properties of, respectively, the
return probability, the spreading of the wave packet, and the average
participation ratio of eigenstates.  

The obtained values of $\delta$ are compatible with the spectral measure being
singular continuous in one dimension and undergoing a transition from singular
to absolute continuous in $d=2,3$ with large subdominant contributions.  The
latter would appear (if such a transition indeed occurs) as a systematic shift
of the value of $\delta(v)$ when the system size increases.  The comparison of
the results for $\delta(v)$ for $m=6$, and $8$ in the two-dimensional systems
(cp.~Fig.~\ref{fig:delta}) shows such a shift of the order of up to $10\%$ near
the speculated transition point $v_c\approx 0.6$ and thus corroborates such
conclusion. A more quantitative characterization of these corrections, such as,
for instance, whether they are of the form (\ref{eq:subdominant}), is beyond
the scope of this work.  The obtained values for $\beta$ are independent of the
dimensionality, which we have linked to the properties of the inverse
participation numbers.  

Even though these results are certainly related to the product structure of the
quasiperiodic systems studied here, it is quite possible that they describe
some features of generic higher-dimensional quasiperiodic systems.  In
particular, we found an exact relation among the exponent $\delta$ and
corresponding exponents $\delta_\alpha$ of the dynamics of the projection onto
subintervals of the spectrum of a wave packet initially localized at a single
site, and compared the dynamics of wave packets in the silver-mean
quasiperiodic tiling constructed from about $2\%$ of the total number of states
near various energies in the band with a study of generalized quasiperiodic
Rauzi tilings in Ref.~[\onlinecite{Triozon03}]. There seems to be a similarity
between the two in so far that wave packets of states near the band edge are
spreading much faster than those made out of the states near the band center.
However, while in the latter work dynamics of these wave packets is,
respectively, ballistic and anomalously diffusive, we find that the dynamics
is, respectively, anomalously diffusive and, for the wave packets made out of
band-center states, slower than any anomalous diffusion.

\begin{acknowledgments}
One of us (V.Z.C.) gratefully acknowledges discussions with M.~Baake and
D.~Lenz.  
\end{acknowledgments}

\end{document}